\documentclass[10pt,a4paper,twoside]{article}
\usepackage{epsfig}
\usepackage{baltlat6}
\usepackage{wrapfig}
\usepackage{lscape}
\begin{document}
\ \
\vspace{0.5mm}
\setcounter{page}{167}
\vspace{8mm}

\titlehead{Baltic Astronomy, vol.\,16, 167--182, 2007}

\titleb{YOUNG STARS IN THE CAMELOPARDALIS DUST AND\\ MOLECULAR CLOUDS.
I. THE CAM OB1 ASSOCIATION}

\begin{authorl}
\authorb{V. Strai\v zys}{} and
\authorb{V. Laugalys}{}
\end{authorl}

\moveright-3.2mm
\vbox{
\begin{addressl}
\addressb{}{Institute of Theoretical Physics and Astronomy, Vilnius
University,\\ Go\v stauto 12, Vilnius LT-01108, Lithuania}
\end{addressl}}

\submitb{Received 2007 April 5; accepted 2007 April 20}

\begin{summary} The distribution of dust and molecular clouds in the
direction of Galactic longitudes 132--158$\degr$ and latitudes
$\pm$\,12\degr\ is investigated.  The maps of dust distribution in the
area were plotted from the following surveys:  the star counts in the
DSS I database by Dobashi et al.  (2005), the survey of the average
infrared color excesses by Froebrich et al.  (2007) and the thermal dust
emission survey at 100~$\mu$m by Schlegel et al.  (1998).  The
distribution of molecular clouds was taken from the whole sky CO survey
by Dame et al.  (2001).  All these surveys show very similar cloud
patterns in the area.  Using the radial velocities of CO, the distances
to separate clouds are estimated.  A revised list of the Cam OB1
association members contains 43 stars and the open cluster NGC 1502.
18 young irregular variable and H$\alpha$ emission stars are
identified in the area.  All this proves that the star forming process
in the Camelopardalis clouds is still in progress.  \end{summary}


\resthead{Young stars in the Camelopardalis dust and molecular clouds.
I.}{V. Strai\v zys, V. Laugalys}

\sectionb{1}{INTRODUCTION}

In the Camelopardalis segment of the Milky Way optical, infrared and
radio observations reveal the presence of numerous dust and molecular
clouds of various densities.  Schlegel et al.  (1998) have mapped dust
emission at 100~$\mu$m from the IRAS and COBE/DIRBE satellite
observations.  Dobashi et al.  (2005) for the evaluation of interstellar
extinction have used a star-count technique in the optical database of
Digitized Sky Survey I (DSS I, red).  Froebrich et al.  (2007) have
presented a map of interstellar extinction in the Galactic anticenter
direction based on the distribution of average color-excesses estimated
in the {\it J,H},$K_s$ system from the 2MASS survey.

Surveys of the molecular CO radio emission in Camelopardalis and the
nearby regions were published by Dame et al.  (1987, 2001), Digel et al.
(1996) and Brunt et al.  (2003).  Digel et al. show that molecular
clouds in the investigated direction are formed by three layers with
different radial velocities:  (1) the local layer (hereafter the Gould
Belt layer) with velocities between --5 and +10 km/s, \hbox{(2) the Cam
OB1} layer with velocities between --5 and --20 km/s, and (3) the
Perseus arm with velocities between --30 and --60 km/s.  Radial
velocities, transformed to distances from the Sun applying the Galactic
rotation curve, give the following mean distances of the molecular
layers in the Local arm:  $\sim$\,200 pc for the Gould Belt layer and
$\sim$\,800 pc for the Cam OB1 association layer.

The interstellar extinction in four Cam areas was investigated in the
seven-color {\it Vilnius} photometric system (Zdanavi\v cius et al.
1996, 2001, 2002a,b, 2005a,b).  A common property of all areas is the
extinction rise after 120--150 pc reaching 1.5--2.5 mag at 1 kpc.
At larger distances the extinction values are 1.5--2.0 mag in relatively
transparent areas and 2.0--3.0 mag in the direction of dark clouds.
However, the heavily reddened stars in these areas were too faint to be
measured; consequently, there is a selection effect present. This means
that the stars with large extinction ($A_V > 4$ mag) were
not reached in these studies.

Here we will analyze the area in the ranges of Galactic longitudes
132--158$\degr$ and latitudes $\pm$\,12$\degr$.  This area covers all
the Milky Way segment in Camelopardalis and also includes the edges of
Cassiopeia, Perseus and Auriga constellations.  Our attention will be
concentrated on the distribution of dust and molecular clouds and young
objects related to star forming in the Local spiral arm.

  The most prominent object, which has determined
boundaries of the investigated area, is the Cam OB1 association.  It is
located at the edge of the Local arm, outside the traditional Gould Belt
(see Elias et al. 2006).  Racine (1968) suggests that the two
A-supergiants, HD 21291 and HD 21389, illuminating the reflection
nebulae vdB\,14 and vdB\,15, and the three other young stars in the
vicinity, form a separate association Cam R1.  However, the necessity
to introduce a new aggregate of stars is doubtful since all
these stars belong to the central part of the Cam OB1 association.

A more complete review of the investigations in the Camelopardalis area
will be published in the Handbook of Star Forming Regions (Strai\v zys
\& Laugalys 2007).


\begin{figure}[!t]
\centerline{\psfig{figure=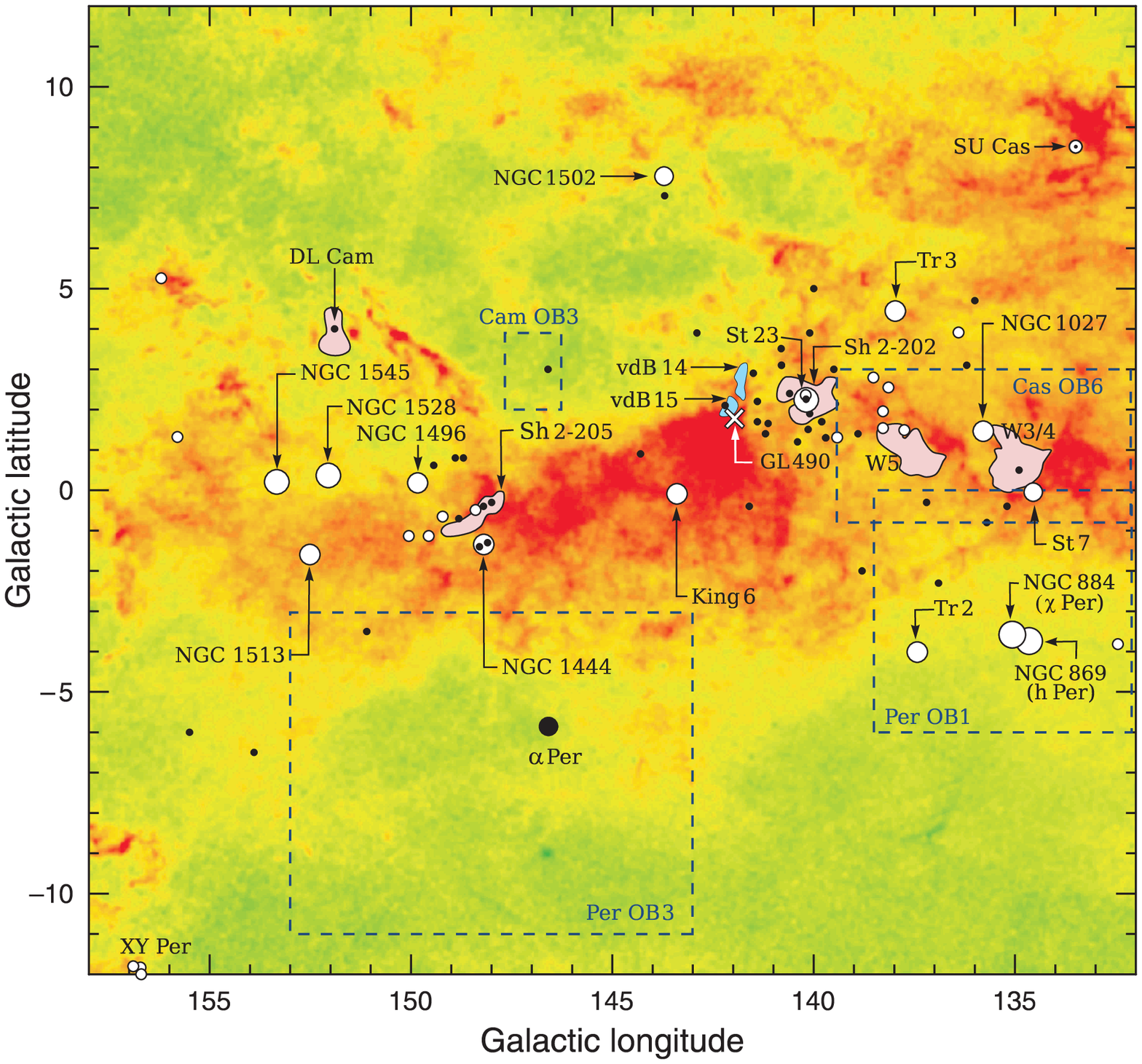,width=124mm,angle=0,clip=true}}
\vspace{0mm}
\captionb{1}{Dust clouds in Camelopardalis and the nearby regions from
Dobashi et al.  (2005) plotted together with the known young stellar
objects, clusters and nebulae.  Black dots designate Cam OB1
association
 members, open circles designate young irregular variables and
H$\alpha$ emission stars.  The YSO GL\,490 is shown as a white
cross.  Positions of 12 open clusters belonging to the Local arm
and of the double cluster h+$\chi$ Per are also shown.
The bright rosy patches designate the emission nebulae W3/4, W5,
Sh2-202, Sh2-205 and DL Cam, the two small blue patches are
reflection nebulae vdB 14 and 15.  The four rectangles limit the areas
of the associations Cas OB6 and Per OB1, located in the Perseus arm,
Cam OB3 -- in the Outer arm and Per OB3 -- in the foreground of Cam
OB1.}
\end{figure}

\sectionb{2}{THE CAM OB1 ASSOCIATION STARS}

We will start from the revision of the suspected members of the Cam OB1
association.  The main criterion of membership in the Cam OB1
association is distances estimated from the MK spectral type and $B$,
$V$ photometry.  In this case the main source of the distance errors are
luminosity classes.  An error of one MK luminosity class for stars of
early B subclasses corresponds to the error of about $\pm$\,1 mag, and
this results in the distance errors of about 50\,\%.  Thus, we
considered a star to be a member of the association if its calculated
distance was between 500 and 1500 pc.  Some potential association stars
possess spectral peculiarities (emission, duplicity, variability etc.),
and their distances are estimated with lower accuracy.  An additional
criterion was the radial velocity (if available):  the stars with
$v_{\rm r}$\,$<$\,--30 km/s were considered as non-members.

Applying these criteria, from the Humphreys \& McElroy (1984) list of 53
stars we excluded 11 stars probably belonging to the Per OB1 and Cas OB6
associations.  Two stars of the cluster NGC 1502 were also excluded,
since we treat NGC 1502 as a separate part of the association.  Three
B0--B2 stars at the longitudes 152--156\degr\ having distances close to
1 kpc were added to the list.  Our revised list presented in Table 1
contains 43 stars:  two O8.5--O9 stars, 35 B0--B3 stars and six A, G and
K supergiants.

However, the membership of BD+62 480, HD 20041, HDE 237204, HD 24094,
Hiltner 404 and HD 25517 is doubtful since the calculated distances
place these stars beyond the Local arm.  Their distances become smaller
than 1500 pc if we reduce their luminosities by about 0.5--0.7 mag.
This means that luminosity V stars should be accepted as belonging to
the zero-age main sequence.

The distribution of the Cam OB1 stars in Galactic coordinates is shown
in Figure 1. The stars concentrate in the three groups.
Group Cam OB1-A, with a

\begin{table}[!t]
\begin{center}
\vbox{\small\tabcolsep=5pt
\parbox[c]{124mm}{\baselineskip=9pt
{\bf\ \ Table 1.}{\small\  Members of the Cam OB1 association
(without NGC 1502 stars).\lstrut}}
\begin{tabular}{rllcrrcccc}
\tablerule
HD, BD, & ~~Var.  & ~Sp  & $\ell$ & $b$~~ & $V$~~ & $B$--$V$ & $J$ & $H$ & $K_s$ \\
Hiltner~~ &       &     &  deg   & deg & mag &  mag  &  mag  &  mag  &  mag    \\
\tablerule
  +60 503  &            &   B1.5\,V     &     134.92  &     0.46   &     9.95  &  0.66 &  8.47 &  8.31 &  8.23 \\
  Hil 322  &  NSV 849   &   B0.5\,V     &     135.20  &   --0.42   &    10.52  &  1.08 &  9.24 &  8.98 &  8.85 \\
  +58 488  &            &   B0.5\,V     &     135.70  &   --0.78   &     9.85  &  0.68 &  8.39 &  8.25 &  8.16 \\
    17958  &  NSV 985   &   K3\,Ib      &     135.96  &     4.67   &     6.28  &  1.90 &   --  &   --  &   --  \\
  +62 480\rlap{*} &  TX Cas    &   B0\,V  &   136.26  &     3.08   &     9.00  &  0.34 &  8.71 &  8.66 &  8.63 \\
    16264  &            &   B1\,Ve      &     136.95  &   --2.33   &     9.25  &  0.44 &  8.30 &  8.17 &  8.08 \\
    17114\rlap{*} &  V792 Cas  &   B1\,V  &   137.26  &   --0.32   &     9.17  &  0.50 &  8.06 &  7.95 &  7.90 \\
    18877  &            &   B7\,II-III  &     138.89  &     1.36   &     8.33  &  0.18 &  7.99 &  7.98 &  7.96 \\
    19968  &            &   B3\,III     &     139.46  &     2.97   &     7.55  &  0.08 &  7.35 &  7.39 &  7.38 \\
    19441  &            &   B3\,III     &     139.72  &     1.26   &     7.87  &  0.37 &  7.19 &  7.12 &  7.13 \\
    19644\rlap{*} &  V368 Cas  &   B3\,III &  139.77  &     1.75   &     8.20  &  0.25 &  7.80 &  7.81 &  7.72 \\
    21212\rlap{*} &  CR Cam    &   B2\,V:e &  140.01  &     4.96   &     8.13  &  0.59 &  6.69 &  6.49 &  6.26 \\
   237090\rlap{*} &  V803 Cas  &   B1\,Ve &   140.05  &     1.91   &     8.86  &  0.53 &  7.78 &  7.68 &  7.57 \\
    20798  &            &   B2\,IV      &     140.11  &     3.87   &     8.37  &  0.25 &  7.80 &  7.81 &  7.80 \\
    19820\rlap{*} &  CC Cas    &   O8.5\,III & 140.12 &     1.54   &     7.10  &  0.51 &  6.07 &  6.00 &  5.94 \\
    20134\rlap{*} &  NSV 1087  &   B2\,IVe &  140.16  &     2.16   &     7.47  &  0.10 &  7.19 &  7.19 &  7.19 \\
  +58 578  &            &   B3\,III     &     140.44  &     1.20   &     9.68  &  0.49 &  8.94 &  8.96 &  8.86 \\
    20508  &  NSV 1101  &   B1.5\,IV    &     140.60  &     2.39   &     8.23  &  0.47 &  7.37 &  7.30 &  7.23 \\
   +60 682 &            &   B3\,III     &     140.75  &     3.47   &     9.65  &  0.66 &  8.27 &  8.16 &  8.12 \\
    20898  &  NSV 1129  &   B1\,IV      &     140.80  &     3.06   &     7.92  &  0.45 &  6.97 &  6.89 &  6.85 \\
    20547\rlap{*} &            &   B2.5\,III & 141.09 &     1.68   &     8.18  &  0.37 &  7.58 &  7.56 &  7.53 \\
   237121  &            &   B0.5\,V     &     141.21  &     1.39   &     8.94  &  0.46 &  8.02 &  7.96 &  7.87 \\
   237130  &            &   B3\,III     &     141.43  &     1.71   &     9.15  &  0.34 &  8.44 &  8.41 &  8.36 \\
    20959  &            &   B3\,III     &     141.44  &     2.23   &     8.00  &  0.27 &  7.40 &  7.37 &  7.38 \\
    21291\rlap{*} &  CS Cam    &   B9\,Ia &   141.50  &     2.88   &     4.20  &  0.41 &   --  &    -- &    -- \\
    20041  &            &   A0\,Ia      &     141.57  &   --0.41   &     5.78  &  0.74 &   --  &   --  &   --  \\
    21389\rlap{*} &  CE Cam    &   A0\,Ia  &  142.19  &     2.06   &     4.53  &  0.56 &   --  &   --  &   --  \\
    22764\rlap{*} &            &   K4\,Ib  &  142.87  &     3.90   &     5.71  &  1.71 &   --  &   --  &   --  \\
    25443\rlap{*} &            &   B0.5\,III & 143.68 &     7.35   &     6.74  &  0.33 &  6.11 &  6.14 &  6.11 \\
    22253  &            &   B0.5\,III   &     144.28  &     0.92   &     6.53  &  0.33 &  5.84 &  5.82 &  5.79 \\
   237204  &            &   B0.5\,V     &     146.60  &     2.98   &     9.14  &  0.31 &  8.58 &  8.58 &  8.55 \\
    24094\rlap{*} &  CY Cam    &   B1\,III &  147.98  &   --0.33   &     8.30  &  0.40 &  7.45 &  7.38 &  7.35 \\
    23675\rlap{*} &  NSV 1333  &   B0.5\,III & 148.10 &   --1.29   &     6.72  &  0.42 &  5.87 &  5.79 &  5.78 \\
  Hil 404  &            &   B1\,V       &     148.28  &   --0.42   &    11.09  &  0.75 &  9.18 &  8.93 &  8.80 \\
    23800\rlap{*} &            &   B1\,IV   & 148.32  &   --1.34   &     6.94  &  0.32 &  6.16 &  6.12 &  6.00 \\
    25056  &            &   G0\,Ib      &     148.72  &     0.80   &     7.03  &  1.20 &   --  &  4.42 &  4.20 \\
    24431\rlap{*} &            &   O9\,III  & 148.84  &   --0.71   &     6.72  &  0.38 &  5.92 &  5.84 &  5.82 \\
   232874  &            &   B0.5\,V     &     148.87  &     0.78   &     8.86  &  0.41 &  7.97 &  7.92 &  7.81 \\
    25348\rlap{*} &  DE Cam    &   B1\,Ve &   149.38  &     0.67   &     8.33  &  0.21 &  7.34 &  7.16 &  6.94 \\
    24432  &            &   B3\,II      &     151.12  &   --3.50   &     6.82  &  0.55 &  5.58 &  5.47 &  5.38 \\
    28446\rlap{*} &  DL Cam    &   B0\,III &  151.91  &     3.95   &     5.77  &  0.16 &  5.43 &  5.46 &  5.45 \\
    24560\rlap{*} &  V581 Per  &   B2\,III &  153.89  &  --6.53 & 8.00\rlap{:} &  0.22 &  6.94 &  6.81 &  6.63 \\
    25517  &            &   B1\,V       &     155.52  &   --6.00   &     9.27  &  0.25 &  8.75 &  8.76 &  8.73 \\
\tablerule
\end{tabular}
}
\end{center}
\end{table}
\vskip3mm

\newpage

\noindent {\bf Notes:}
\vskip2mm

\noindent TX Cas -- eclipsing binary of EB type, $P$ = 2.9 d;

\noindent V792 Cas -- variable of $\beta$ Cep type;

\noindent V368 Cas -- eclipsing binary of EA type, $P$ = 4.45 d;

\noindent CR Cam -- variable of Be type; possible Gould Belt star at 430 pc;

\noindent CC Cas -- eclipsing binary of EB type, $P$ = 3.4 d; ionizes
the Sh2-202 nebula

\noindent V803 Cas -- variable of Be type;

\noindent HD 20134 -- within the doubtful cluster Stock 23;

\noindent HD 20547 -- binary;

\noindent CS Cam -- variable of $\alpha$ Cyg type, illuminates the
reflection nebula vdB\,14;

\noindent CE Cam -- variable of $\alpha$ Cyg type, illuminates the
reflection nebula vdB\,15;

\noindent HD 22764 -- binary;

\noindent HD 25443 -- near the open cluster NGC 1502;

\noindent CY Cam -- long-period pulsating B-type star;

\noindent HD 23675 -- within the doubtful cluster NGC 1444, multiple system;

\noindent HD 23800 -- within the doubtful cluster NGC 1444;

\noindent HD 24431 -- binary; ionizes the Sh2-205 nebula;

\noindent DE Cam -- variable of Be type;

\noindent DL Cam -- variable of $\beta$ Cep type, triple system, surrounded by
          a H$\alpha$ emission region of 1.5--2\degr\ diameter (see
          Finkbeiner 2003 and http://skyview.gsfc.nasa.gov);

\noindent V581 Per -- variable of $\beta$ Cep type.

\vskip3mm

\noindent diameter of 6\degr, contains almost a half of the association
members, most of them concentrate around the Sh2-202 emission nebula and
the vdB\,14 and vdB\,15 reflection nebulae, with a center at $\ell$ =
140\degr, $b$ = +1.5\degr.  Group Cam OB1-B of similar size, containing
$\sim$\,13 members, concentrates around the emission nebula Sh2-205 and
the dust ring at $\ell$ = 150--151\degr, which will be described in
Section 4. Group Cam OB1-C contains the open cluster NGC 1502 located at
$\ell$, $b$ = 143.7$\degr$, +7.7$\degr$ and one star (HD 25443) nearby.
In the first lists this cluster was considered as a separate association
Cam II; later on, some of its stars were incorporated into the list of
the Cam OB1 members (Ruprecht 1964; Humphreys \& McElroy 1984).

In the area of Group A, a massive young stellar object (or a
Becklin-Neugebauer object) GL\,490 (Snell et al. 1984; Campbell et al.
1986; Alvarez et al. 2004) is located; in Figure 1 it is shown as a
white cross.  It is considered to be a massive protostar in the
gravitational contraction stage, surrounded by a dust and gas envelope,
which gives an extinction $A_V$ of more than 17 mag, estimated from the
absorption in the silicate band at 9.7 $\mu$m.

The distance of the Cam OB1 association can be estimated by taking the
average distance of its members.  For this aim we chose stars with
reliable MK classification, excluding binaries and emission-line stars,
but accepting small-amplitude variables of $\beta$ Cep and $\alpha$ Cyg
types.  The absolute magnitudes, $M_V$, for MK spectral types were taken
from the Strai\v zys (1992) tabulation.  For luminosity class V stars we
took the average absolute magnitude for this luminosity (not for the
ZAMS stars).  The ratio $R=A_V/E_{B-V} = 2.90$ was accepted (Zdanavi\v
cius et al. 2002c).  For 26 selected stars the obtained average distance
is 1010\,$\pm$\,210 pc (standard deviation).  This value is in
reasonable agreement with the earlier determinations by Humphreys (1978,
1.0 kpc), Melnik \& Efremov (1995, 0.98 kpc), Zeeuw et al.  (1999, 0.9
kpc) and Lyder (2001, 0.98 kpc).  At a distance of 1 kpc the diameters
of both groups of Cam OB1 stars are about 100 pc what is a typical size
of other associations.

The same procedure applied to 11 B0--B3 stars in NGC 1502 (for which
{\it BV} photometry is available) gives a distance of 1180\,$\pm$\,160
pc if the average $M_V$ values for luminosity V stars are taken, and
860\,$\pm$\,130 pc if $M_V$ for the ZAMS stars are taken.  The real
distance should be somewhere in between.  Within errors this distance is
close to that of the loose stars of Cam OB1.

Lyder (2001) has attempted to identify Cam OB1 stars of lower masses
(spectral classes B5--A0 V) in the area of our Group A. The spectral,
photometric and radial velocity data have been collected from the
literature.  However, the membership of the selected stars should be
verified by more accurate spectral classification, photometry and radial
velocities.

\sectionb{3}{OPEN CLUSTERS IN THE LOCAL ARM}

Thirteen open clusters belonging to the Local arm are present in the
area.  They are listed in Table 2 and shown in Figure 1. Table 2 gives
their data from the WEBDA database.  Here we describe properties of some
of the clusters located near the outer edge of the Local arm and discuss
their possible relation to the association.  The clusters residing in
other arms or with unknown distances are not considered.

\vskip1mm

\begin{table}[!h]
\begin{center}
\vbox{\small\tabcolsep=6pt
\parbox[c]{120mm}{\baselineskip=10pt
{\normbf\ \ Table 2.}{\small\ Open clusters in the Local spiral arm. The
data are taken from the WEBDA database.\lstrut}}
\begin{tabular}{lcrrrl}
\tablerule
Name & $\ell$ & $b$~ & $d$~ & Age & Notes \\
     &  deg   &  deg & pc & Myr &        \\
\tablerule
Stock 7   &  134.6 &   0.0 &   700  &    16 &  \\
NGC 1027  &  135.8 &  +1.5 &   770  &   160 &  \\
Trumpler 2 & 137.4 &  -4.0 &   650  &   148 &  \\
Trumpler 3 & 138.0 &  +4.5 &   830  &   219 &  \\
Stock 23  &  140.1 &  +2.1 &   380  &    32 & doubtful \\
King 6    &  143.4 & --0.1 &   870  &   219 &          \\
NGC 1502  &  143.7 &  +7.7 &   820  &    11 & belongs to Cam OB1 \\
$\alpha$ Per & 148.0 & --7.0 & 177  &    50 & Per OB3  \\
NGC 1444  &  148.1 & --1.3 &  1200  &    91 & doubtful  \\
NGC 1496  &  149.8 &  +0.2 &  1230  &   631 & \\
NGC 1528  &  152.1 &  +0.3 &   780  &   372 & \\
NGC 1513  &  152.6 & --1.6 &  1320  &   129 & \\
NGC 1545  &  153.4 &  +0.2 &   710  &   282 & \\
\tablerule
\end{tabular}
}
\end{center}
\vskip-2mm
\end{table}

\enlargethispage{3mm}

{\bf NGC 1502}

This cluster is
considered as a part of the Cam OB1 association since it contains 17
B0--B3 stars (Tapia et al. 1991), and both the cluster and association
are located at a similar heliocentric distance.  Two brightest stars of
the cluster, HD 25638 and HD 25639, sometimes are classified as O9 or
O9.5 stars (see the WEBDA database and Maiz-Apell\'aniz et al.
2004). One more star, HD 25443, of spectral type B0.5\,III is located at
a distance of 0.3\degr\ from the cluster center.  In Figure 1 this
star is shown separately.

\vskip2mm

{\bf Stock 23}

This is a poor group of stars, seen in the direction of the Sh2-202
emission nebula which is located in the Local arm at 800 pc from the Sun
(Fich \& Blitz 1984).  Kharchenko et al.  (2005) find that Stock 23 is a
foreground object at a distance of 380 pc.  In the $V$ vs.  $B$--$V$
diagram the stars of this group are so scattered that the cluster
probably is not real.  One of stars in the group of spectral type
B2\,IV, HD 20134, may be a member of Cam OB1.  On a deep CCD picture
obtained by Dean Salman (Internet web site galaxies.com) the group seems
to be surrounded by a blue cloud which might be a reflection nebula but
it is unknown which star illuminates the dust.

\vskip2mm
{\bf Stock 7}

A compact group of stars at the southern edge of the W4 nebula, contains
eight B-type stars. Two brightest stars of spectral type B5\,V are
variables V529 Cas (HD 15238, $\gamma$ Cas type) and V528 Cas (HD 15239,
$\beta$ Cep type). More details are given by Moffat \& Vogt (1973).

\vskip2mm
{\bf NGC 1027}

A medium-age cluster at the 770 pc distance, seen at the edge of the W4
nebula.  The cluster image is dominated by a peculiar metal-deficient F4
star HD 16626 ($V = 6.99$) which probably has no relation with the
cluster.

\vskip2mm
{\bf King 6}

This is a rich cluster with a well-defined main sequence, the earliest
members being of spectral class B5--B7.  It is projected on the dust
cloud Tokyo 942.  Although its heliocentric distance, 870 pc (Ann et
al. 2002), is similar to that of the association, probably they have no
physical or evolutionary relation due to age differences.

\vskip2mm
{\bf NGC 1444}

The cluster is located in the vicinity of the Sh2-205 emission nebula at
a distance of 900 pc (Fich \& Blitz 1984).  Its image is dominated by
two stars -- HD 23675 and HD 23800 -- of 6.7 and 6.9 visual magnitudes
and B0.5\,III and B1\,IV spectral types.  Both of these stars are in the
list of the Cam OB1 members (Table 1).  Pe\~na \& Peniche (1994) place
the cluster at 906 pc but there are some doubts whether it is a real
cluster or just an accidental group of field stars.

\vskip2mm
{\bf NGC 1496, NGC 1528, NGC 1513 and NGC 1545}

These four clusters are seen near the Galactic plane in the direction of
northern Perseus with longitudes 150--153\degr.  Their heliocentric
distances (Table 2) are between 700 and 1300 pc, i.e., they are close to
the outer edge of the Local arm and to the Cam OB1 association.  Other
parameters of the clusters are given in Table 2. Their ages are in
between the Pleiades and Hyades ages, consequently, they are not related
to the Cam OB1 association.


\begin{figure}[!t]
\centerline{\psfig{figure=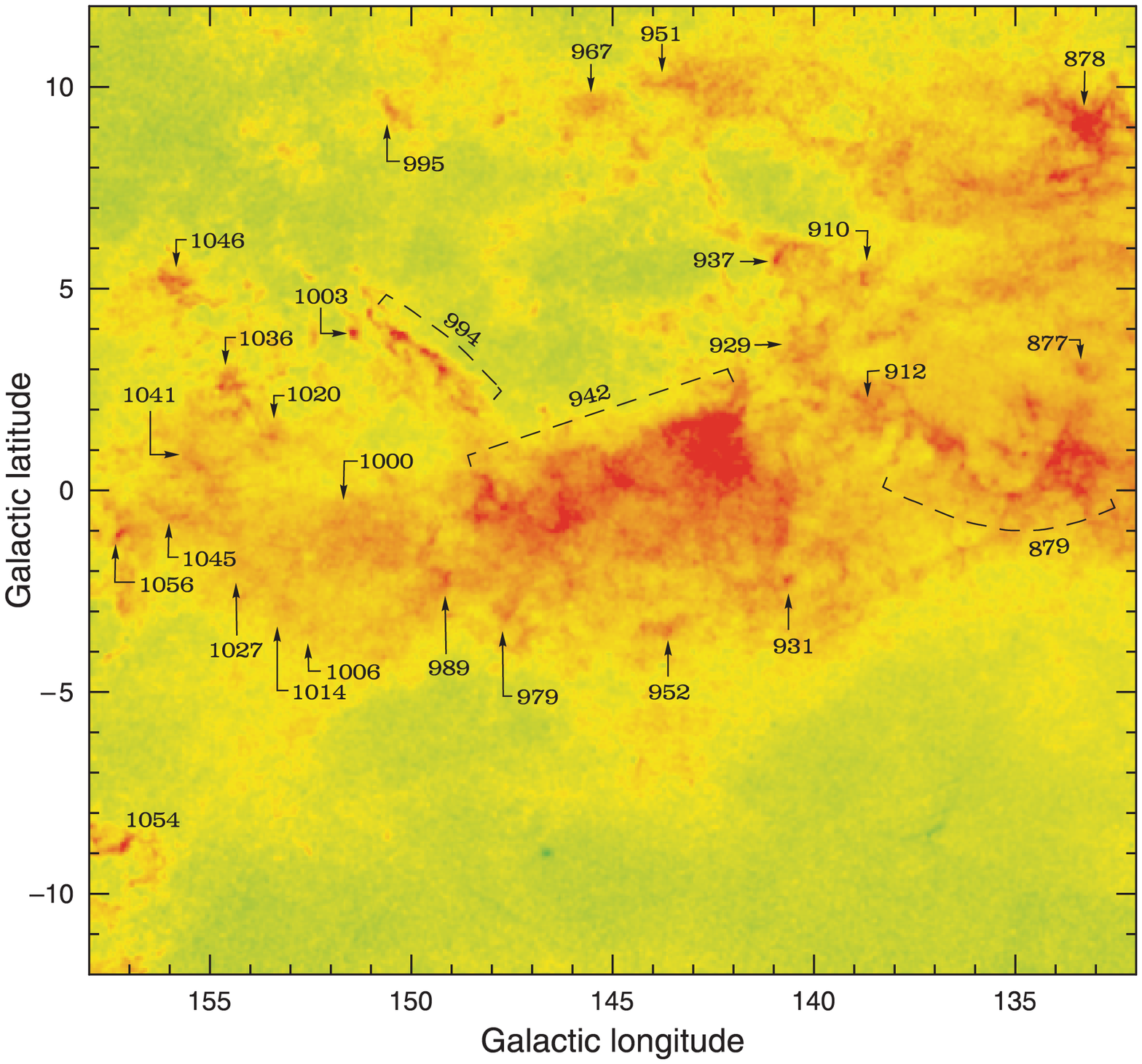,width=124mm,angle=0,clip=true}}
\vspace{0mm}
\captionb{2}{Dust clouds from Dobashi et al.  (2005) with the Tokyo
numbers of the darkest clouds.}
\end{figure}

\sectionb{4}{DUST AND MOLECULAR CLOUDS}

Figure 2 repeats the distribution of dust clouds from Dobashi et al.
(2005) without the other objects plotted.  The darkest clouds are
identified by the Tokyo numbers.  Figure 3 shows a map of interstellar
extinction based on the distribution of average $J$--$H$ and $H$--$K$
color excesses estimated from the 2MASS survey \hbox{(Froebrich} et al.
2007).  Figure 4 shows the distribution of the dust emission at 100
$\mu$m composed from the observations with the IRAS and COBE satellites
(Schlegel et al. 1998).

All these three surveys correspond to the extinction with somewhat
different penetration into space.  The 100 $\mu$m dust emission method
has the deepest penetration corresponding to the entire dust amount or
extinction in the given directions.  The extinctions from the average
color excesses based on {\it J,H,K} photometry also correspond to very
deep penetration, being able to measure extinctions $A_V$ as large as 20
mag or more.  In the Galactic anticenter direction such large
extinctions are met only in a few directions coinciding with the densest
clouds.  The method used by Dobashi et al., based on star counts in the
Palomar DSS, detects all the nearby clouds and the distant clouds which
are not hidden by the foreground extinction.

\begin{figure}[!t]
\centerline{\psfig{figure=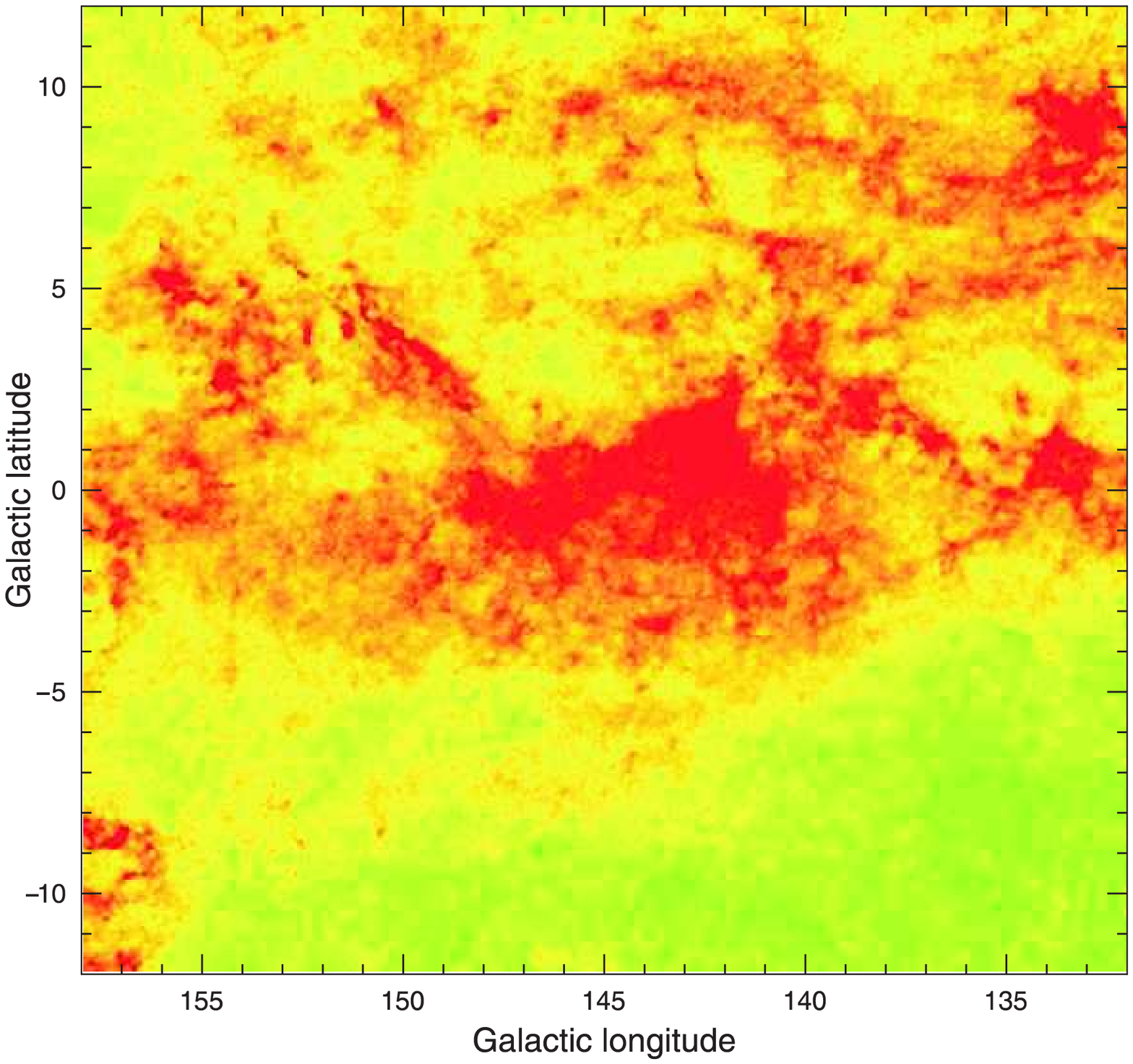,width=124mm,angle=0,clip=true}}
\vspace{0mm}
\captionc{3}{Dust clouds from Froebrich et al. (2007).}
\end{figure}

The distribution of CO molecular clouds taken from Dame et al.  (2001)
is shown in Figure 5. A comparison of the CO and dust distributions in
the area exhibit close resemblance.  Although most of the CO clouds
concentrate mostly between the latitudes --4\degr\ and
+12\degr, some of them are observed even at +24$\degr$ from the Galactic
equator (Heitshausen et al. 1993; Dame et al. 2001).

To find the cloud assignment to different layers, we have made a radial
velocity analysis for the Dame et al.  (2001) data.  The velocity
divisions were the same as in the Digel et al.  (1996) paper.  The
resulting picture is shown in Figure 6 where the clouds of the Gould
Belt layer (150--300 pc from the Sun) are shown in blue, of the Cam OB1
layer (at 800--900 pc) -- in green and in the Perseus arm ($>$\,2 kpc)
-- in red.  The comparison of our Figure 6 with Figures 3--5 of the
Digel et al. survey shows that in the overlapping longitude range
(132--144\degr) the assignment of clouds to different layers coincides.
It is evident that most clouds seen in the area belong to the two layers
of the Local arm.  The input of the Perseus arm is relatively small, and
these distant clouds concentrate close to the Galactic equator.

\begin{figure}[!t]
\centerline{\psfig{figure=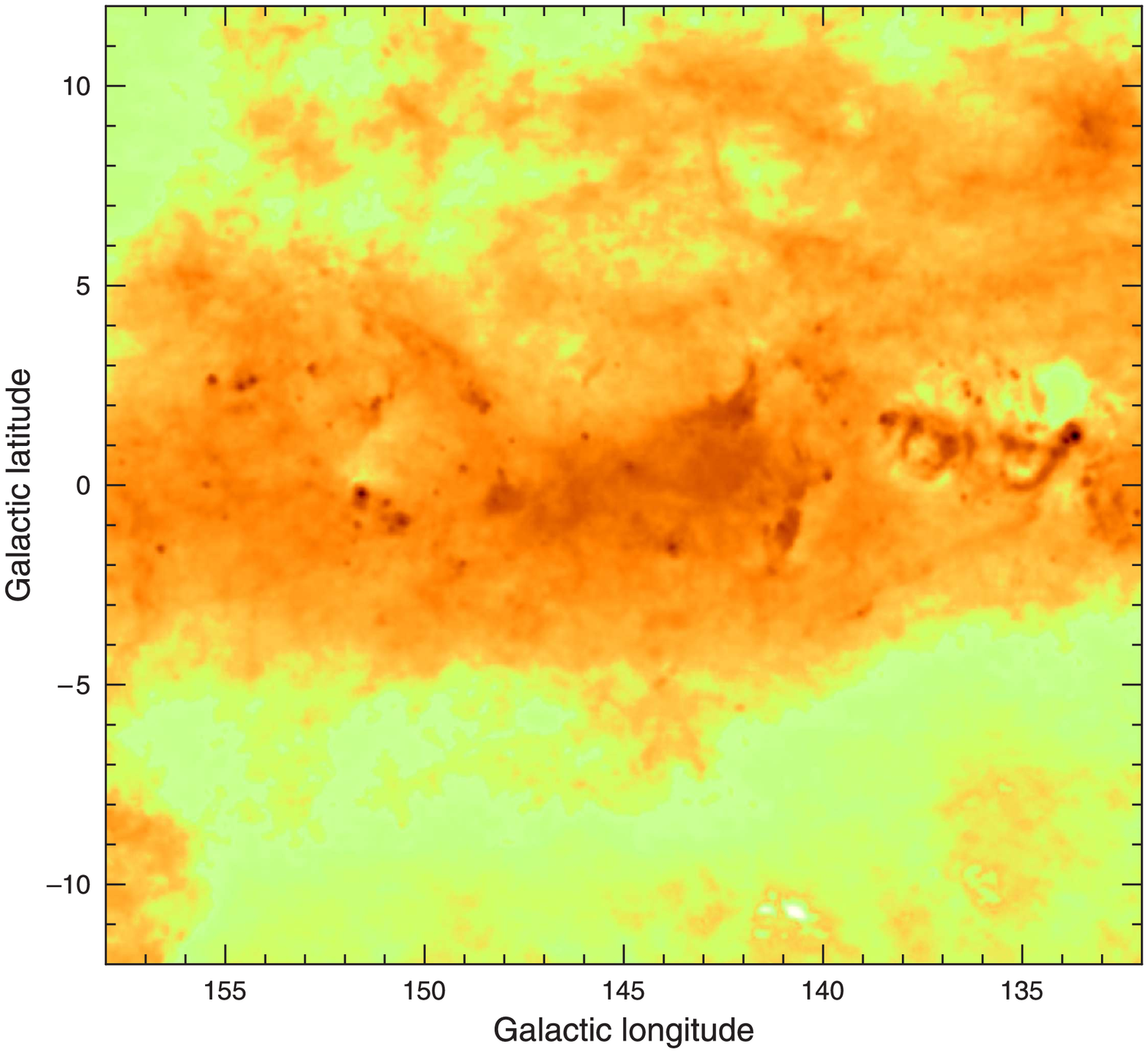,width=124mm,angle=0,clip=true}}
\vspace{0mm}
\captionc{4}{Dust clouds from Schlegel et al. (1998).}
\end{figure}

A comparison of Figures 2--5 shows that all of them exhibit a quite
similar pattern of dust clouds.  This happens because most of the clouds
in this direction are located in the Local and the Perseus arms up to
about a 3 kpc distance.  However, the Schlegel et al. dust map shows
more dense dust clumps located in the Perseus arm and maybe in the Outer
arm.  At shorter wavelengths these clumps are not seen being covered by
the nearer clouds.  Hereafter we will base our comparison of various
objects on the map of Dobashi et al. which better represents dust
distribution in the Local arm.  Here we will discuss the main features
of the dust and molecular cloud pattern by comparing it with the
positions and distances of other objects seen in the same directions and
shown in Figure 1.

A general feature in the investigated area is the absence of clouds with
Galactic latitudes lower than --4\degr.  In the foreground, this region
is occupied by stars of the Per OB3 (or $\alpha$ Per) association
located at a distance of 177 pc (Zeeuw et al. 1999).  The hottest
members of this association are of spectral class B3, and its age is
about 50 Myr. In the left lower corner of our area a few clouds are
seen, they are the northern end of a large cloud structure, which
includes the California Nebula and belongs to the Per OB2 association
system.

\begin{figure}[!t]
\centerline{\psfig{figure=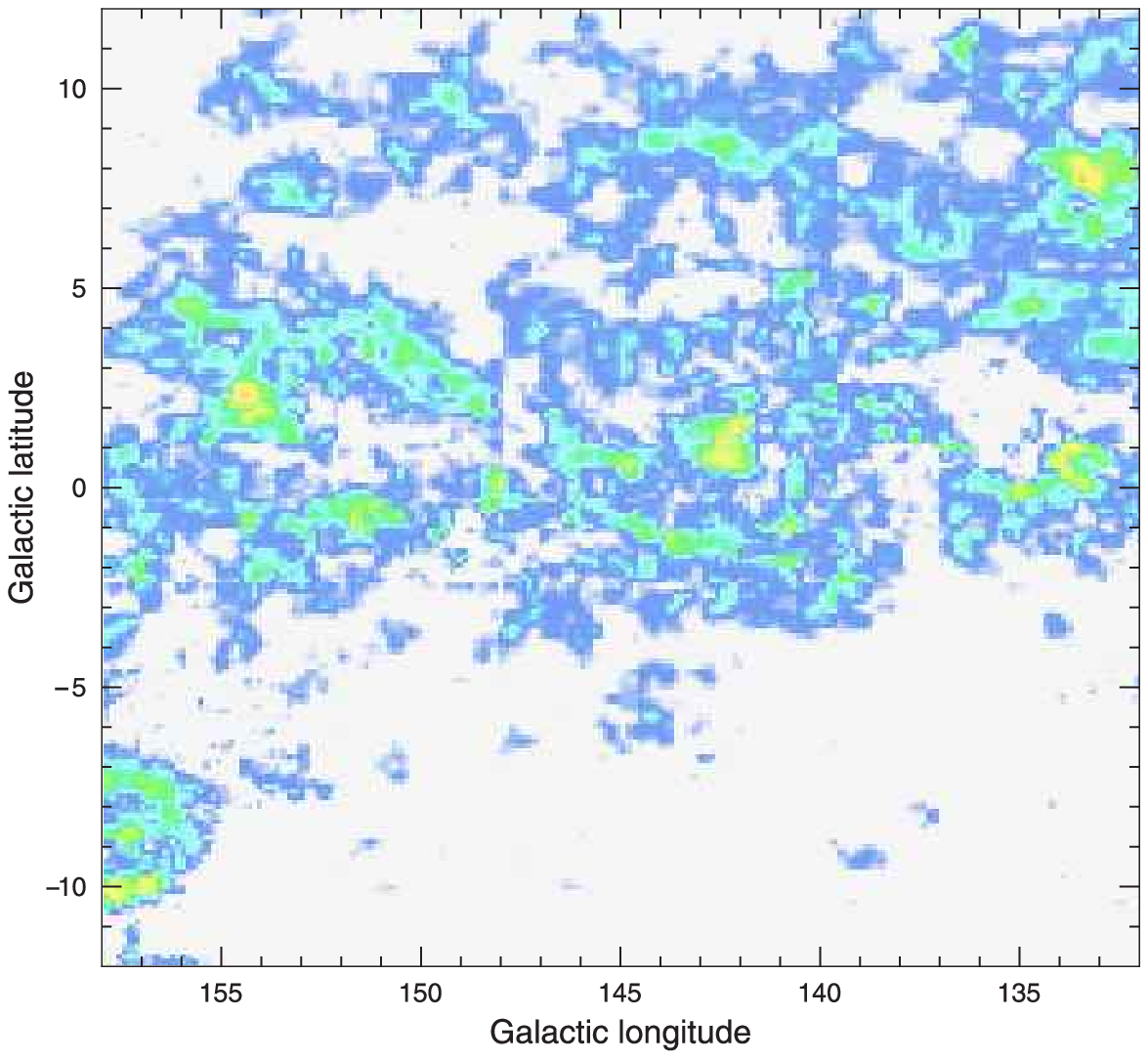,width=124mm,angle=0,clip=true}}
\vspace{0mm}
\captionc{5}{CO clouds from Dame et al. (2001).}
\end{figure}

A further story about the dust and molecular clouds we will start from
the right edge of the area.  Here we find the associations Cas OB6 and
Per OB1 shown in Figure 1 as broken rectangles which are located at a
distance of 2.0--2.2 kpc and belong to the Perseus spiral arm.  The
clouds related to the W3, W4 and W5 nebulae (the upper part of Tokyo 879
extending from 133\degr\ to 138\degr) belong to the same star-forming
region as the Cas OB6 association.  The most prominent object of the Per
OB1 association is the double cluster h+$\chi$ Persei.

The Cam OB1 association and the related to it cloud system partly
overlap in projection both the Cas OB6 and Per OB1 associations.  The
lower part of the Tokyo 879 P1 clump with negative latitudes in the
longitude range 134--137\degr\ belongs to the Local arm, mostly to the
Cam OB1 layer.  The clouds at the positive latitudes 1--6.5\degr\ in the
longitude range 138--144\degr\ (Tokyo 912, 929, 937) exclusively
belong to the Cam OB1 layer.  Here is the largest concentration of the
Cam OB1 member stars and the Sh2-202 emission nebula.

One more cloud, Tokyo 878, located at the upper right corner of the
area, deserves a special discussion.  Figure 6 shows that the lower part
of this cloud belongs to the Gould Belt layer while its upper part to
the Cam OB1 layer.  The lower part of the clouds is related to a group
of B8--A3 stars and the cepheid SU Cas described by Turner \& Evans
(1984).  These stars illuminate the nearby dust clouds forming some
small reflection nebulae.  The distance to this group is about 260 pc,
i.e., the clouds really belong to the Gould Belt.

\begin{figure}[!t]
\centerline{\psfig{figure=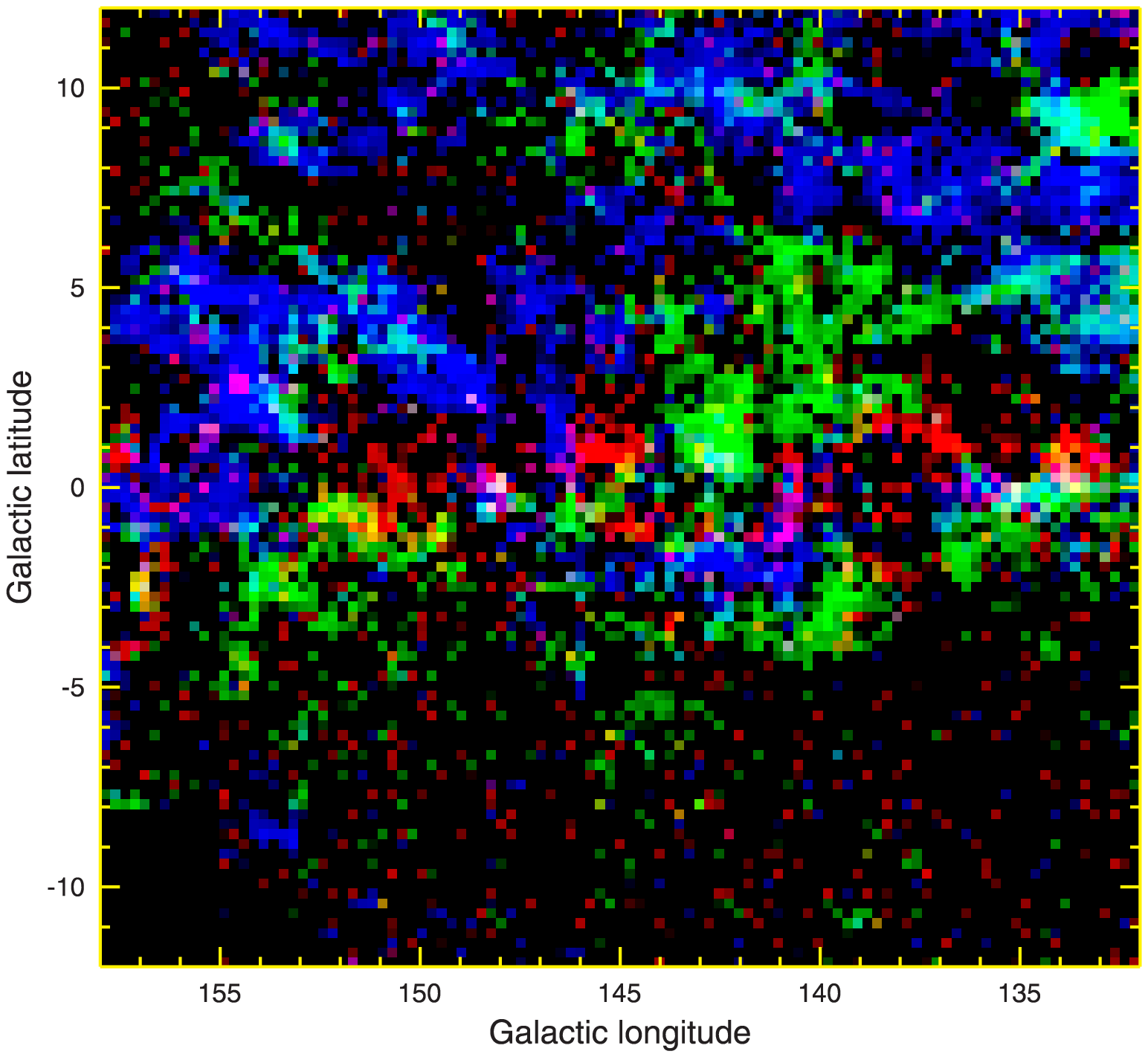,width=124mm,angle=0,clip=true}}
\vspace{0mm}
\captionb{6}{CO clouds from Dame et al. (2001) belonging to different
layers. Clouds of the Gould Belt
layer are shown in blue, of the Cam OB1 layer in green and of the
Perseus arm in red.}
\end{figure}

Let us return to the vicinity of the Galactic equator.  The densest
dust and molecular cloud in the area is Tokyo 942.  In the Dobashi et
al. catalog it is split into 20 condensations numerated from P1 to P20
extending from 142\degr\ to 148\degr\ along the equator.  These
condensations belong to different cloud layers.  For example, the
densest clumps P1, P2 and P3 at 142--144\degr, 0--2\degr\ belong to the
Cam OB1 layer, clump P6 at 145\degr, +0.5\degr\ -- to the Perseus arm,
clump P4 at 146\degr, --0.5\degr\ and clumps P7--P8 at 148\degr,
--0.5\degr\ -- again to the Cam OB1 layer.  The clouds of the Gould Belt
layer appear in many places of the 942 cloud but they are of lower
density.  Among them the largest feature is an oblong cloud running
parallel to the equator between 140--145\degr\ at $b$ = --2\degr.

Farther left dust clouds configure a loop described by Schwartz (1987)
with a diameter of $\sim$\,4$\degr$ and centered at 150$\degr$,
+1$\degr$.  However, this loop is not a real feature -- its clouds
belong to quite different dust layers.  A more realistic loop (or ring)
of ellipsoidal form (7\,$\times$\,5\degr) can be composed of the
following Tokyo clouds:  942 (P7, P8), 994, 1020, 1000 and 989.  Figure
6 shows that all these clouds contain green components of the Cam OB1
layer.  The ring is much better defined in its upper part:  the chain of
small dark clouds and filaments was already known to Barnard (1927).
The ring may be seen both in the dust and the CO maps.  Near the top of
the ring the triple star HD 28446 (DL Cam) with its H$\alpha$ emission
region is located.  A third version of the ring is the circle with a
diameter of $\sim$\,8$\degr$ centered at 152$\degr$, +0.5$\degr$ at the
open cluster NGC 1528.  This ring includes the following Tokyo clouds:
942 (P7, P8), 994, 1003, 1036, 1041, 1027, 1014, 1006, 989 which,
however, belong to two different layers.


\begin{wrapfigure}[29]{r}[0pt]{65mm}
\vspace{-1mm}
\psfig{figure=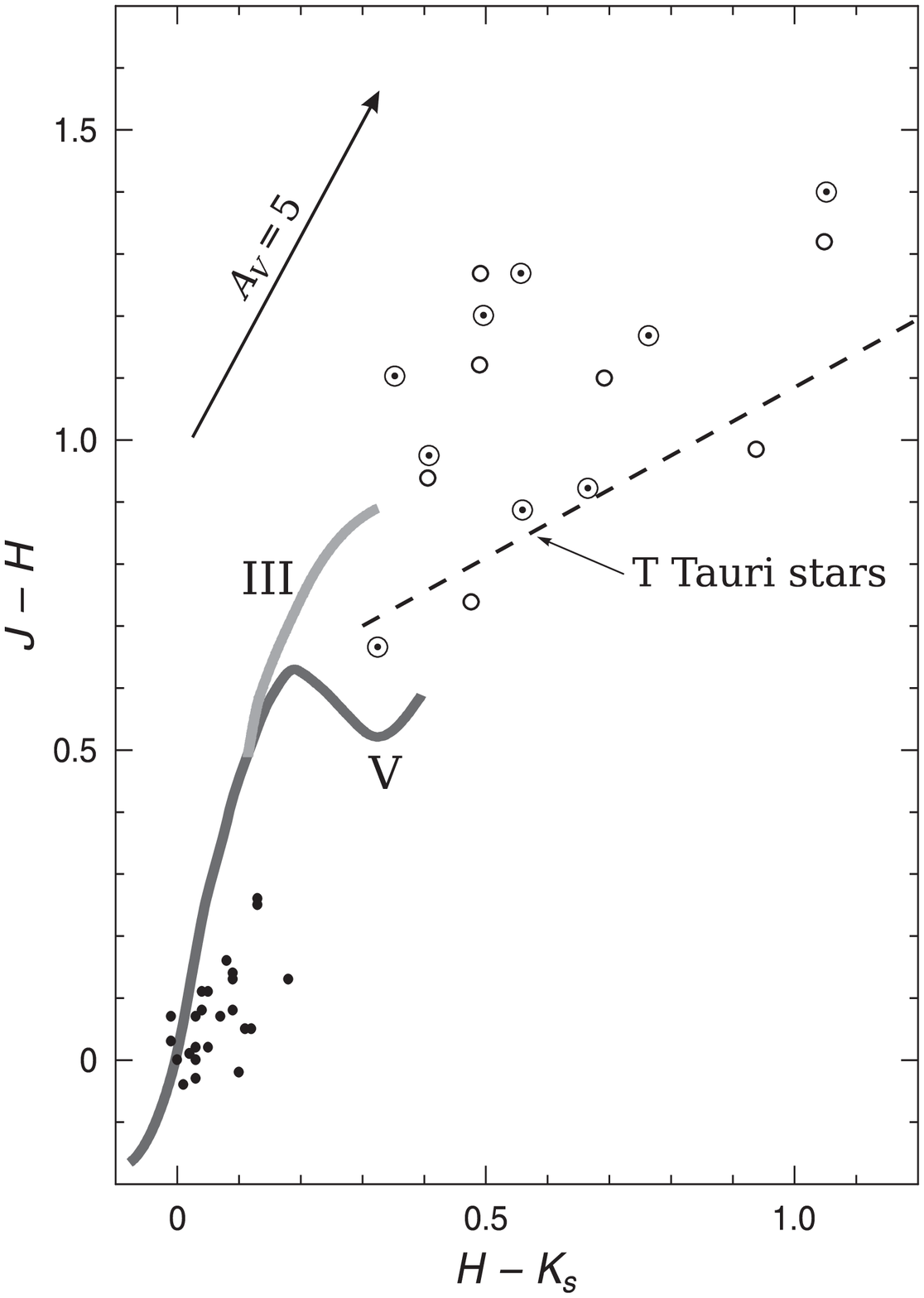,width=64mm,angle=0,clip=}
\vspace{-1mm}
\captionb{7}{\hstrut The $J$--$H$ vs.  $H$--$K_s$ diagram for young
stars in the Camelopardalis area.  Dots designate the O--B stars of the
Cam OB1 association, open circles designate the seven stars with
emission in H$\alpha$, circles with central dots designate nine
irregular variable stars of types IN and IS.  The intrinsic line of T
Tauri stars and interstellar reddening line are also shown.}
\end{wrapfigure}

Most of the clouds to the left and up from the described rings belong to
the Gould Belt layer.  The same is true for the chain of clouds
extending from SU Cas in the right upper corner of the area towards the
cluster NGC 1502 (see Figure 1).  However, Figure 6 shows that in many
places small inserts of clouds from the Cam OB1 layer are seen.

\sectionb{5}{YOUNG STARS OF LOWER\\ MASSES}

In this section we will describe young stars of spectral types from A to
M. Not all of them are really `low-mass stars'; this term usually
applies only to the pre-main-sequence stars of spectral types G, K and M
(T Tauri and post T Tauri stars).  However, spectral types (and masses)
of faint stars with H$\alpha$ emission are mostly unknown.  Therefore we
will use the term `low-mass' stars for the stars cooler than of spectral
type B.

The data on young low-mass stars in the Camelopardalis area are quite
scarce.  The Herbig \& Bell (1988) catalog of emission-line stars of the
Orion population lists only one nebulous H$\alpha$ emission star, IRAS
03134+5958 ($V$\,$\sim$\,14), located near the questionable open cluster
Stock 23, projected on the Sh2-202 nebula.  Other four stars (including
two components of XY Per) from the Herbig \& Bell catalog, falling into
our area, are located in a dark cloud in Perseus, north of the
California Nebula.

More young low-mass stars in Camelopardalis have been found by Gahm
(1990).  In low-dispersion objective-prism spectra, obtained with the
Schmidt telescope of the Stockholm Observatory, he found 12 stars with
emission in H$\alpha$ in the vicinity of the Sh2-205 nebula.  We
identified these stars in the 2MASS catalog and found that in the
$J$--$H$ vs.  $H$--$K_s$ diagram (Figure 7) four of them are located
above or near the intrinsic line of T Tauri stars (Rydgren \& Vrba
1981; Meyer et al. 1997).  The position of the above-mentioned
H$\alpha$ emission star IRAS 03134+5958 in the $J$--$H$ vs.  $H$--$K_s$
diagram also confirms its relation to T Tauri-type objects.  Two more
H$\alpha$ emission stars satisfying the same criterion were found in
the Kohoutek \& Wehmeyer (1997) catalog.  Nine irregular variables of
types IN and IS, selected from the General Catalogue of Variable Stars
(GCVS, Samus et al. 2004), also lie in the same region.  These
variables and the H$\alpha$ emission stars are listed in Table 3 and
plotted in Figure 1 as small open circles.  It is obvious that these
stars are located mostly in dust and molecular clouds, and this also
confirms their possible genetic relationship.  Some of
Table 3 stars can be located in the Perseus arm.

More young objects in the area are described in the literature.  Clemens
\& Barvainis (1988) list seven globulae in our area.  Two of them,
CB\,17 (LDN 1389) and CB\,26 (LDN 1439) are associated with the IRAS
sources identified as protostellar cores (Launhardt \& Henning 1997;
Launhardt \& Sargent 2001).  A HH-like flow and the associated infrared
star are found in a small cloud LDN 1415 (Stecklum et al. 2007), close
to the chain of clouds described in Section 4. H$\alpha$ emission stars
in the Perseus arm H\,II regions W3/4/5 are quite common:  Ogura et al.
(2002) have found there tens of faint emission objects which may be T
Tauri or Herbig Ae/Be stars.

\vskip2mm


\begin{table}[!ht]
\begin{center}
\vbox{\small\tabcolsep=3.5pt
\parbox[c]{110mm}{\baselineskip=9pt
{\smallbf\ \ Table 3.}{\small\ Young irregular variable stars and
H$\alpha$ emission stars. Magnitude ranges for variables are
 from GCVS, blue and red magnitudes for
H$\alpha$ emission stars are from DSS\,I, and {\it J, H, $K_s$}
magnitudes are from 2MASS.\lstrut}}
\begin{tabular}{lcrlrrrl}
\tablerule
Name & $\ell$ & $b$~~ & Magnitude & $J$~~ & $H$~ & $K_s$~ & Notes
\\[-2pt]
     &  deg   &  deg~ & ~~~range   & mag & mag & mag   &       \\
\tablerule
\noalign{\vskip1mm}
\multicolumn{8}{l}{Variables}\\
\noalign{\vskip1mm}
CF Per    &  132.45 &  --3.81 &    10.5--12.5\,$v$ &    5.30 &    4.40 &   --~~      &   ISB, M6 \\
V519 Cas  &  136.40 &   +3.91 &    13.8--15.1\,$p$ &    5.55 &  4.45 &   4.10   & ISB:       \\
LW Cas    &  137.75 &   +1.49 &    15.4--17.1\,$p$ &   12.49 &  11.09 & 10.03 &  INA, A0  \\
V506 Cas  &  138.14 &   +2.55 &    15.6--17.0\,$r$ &   7.46 &   6.19 &   5.64 &  ISB:        \\
V505 Cas  &  138.28 &   +1.96 &    12.9--14.8\,$p$ &   5.90 &   4.94 &   4.53 &  IS:         \\
V508 Cas  &  138.52 &   +2.79 &    15.2--16.4\,$r$ &   7.81 &   6.61 &   6.11 &  ISB         \\
OS Per    &  155.80 &   +1.32 &    16.0--17.5\,$p$ &  10.44 &   9.77 &   9.45 &  IS:         \\
V347 Aur  &  156.20 &   +5.26 &    13.1--16.0\,$p$ &   9.99 &   8.82 &   8.06 & INT, M2\,Ve(T)     \\
LkH$\alpha$\,272  &  156.74 &  --12.00 & 14.0\,$v$,\,var &  10.40 &   9.48 & 8.82 &  INT:, K6e     \\
LkH$\alpha$\,273 &  156.77 &  --11.97 & 15.3\,$v$,\,var &  11.94 &  11.06 &  10.50 &  INT, K7/M0e \\
XY Per    &  156.80 &  --11.90 &   9.4--10.6\,$v$ &   7.65 &   6.92 &   6.09 &  INA, A2e, SB       \\
\noalign{\vskip1mm}
\multicolumn{8}{l}{H$\alpha$ emission stars}\\
\noalign{\vskip1mm}
KW97 14-24    & 138.28 &   +1.54 & 14.7\,$b$, 13.4\,$r$ & 11.79 &  10.80 & 9.87 &   ~~--  \\
KW97 14-52    & 139.42 &   +1.31 & 17.4\,$b$, 13.6\,$r$  & 8.92 & 7.65 & 7.16 &   ~~--  \\
IRAS 03134+5958 & 140.16 &   +2.27 & 10.8\,$b$, 9.3\,$r$   &  9.64 &  8.32 &  7.27 &  H$\alpha$, envelope \\
Gahm 25      & 148.40 &  --0.49 & 17.8\,$b$, 14.9\,$r$  & 10.67 &  9.55 &  9.06 &   ~~--  \\
Gahm 23      & 149.24 &  --0.65 & 16.9\,$b$, 14.3\,$r$  & 10.88 &  9.94 &  9.53 &   ~~--  \\
Gahm 22      & 149.56 &  --0.88 & 18.8\,$b$, 14.2\,$r$  & 11.28 & 10.54 & 10.06 &   ~~--  \\
Gahm 21      & 150.06 &  --1.14 & 16.4\,$b$, 13.8\,$r$  & 12.26 & 11.16 & 10.47 &   ~~--  \\
\noalign{\vskip0.5mm}
\tablerule
\end{tabular}
}
\end{center}
\vspace{-5mm}
\end{table}

\enlargethispage{5mm}

\sectionb{6}{CONCLUSIONS}

On the ground of optical, infrared and radio observations we give the
dust and molecular cloud pattern in Camelopardalis and the nearby areas
of Cassiopeia, Perseus and Auriga (Galactic longitudes from 132\degr\ to
158\degr).  The Local spiral arm in this direction contains many dust
and molecular clouds of high density.  Using the CO radial velocities,
the clouds are divided into the Gould Belt layer at the 150--300 pc
distance and the Cam OB1 association layer at the 800--900 pc distance.
However, in some directions this division may be ambiguous since the
dust distribution across the Local arm can be almost continuous.  The
largest density is observed in the 800--900 pc layer where some clouds
(like Tokyo 942) reach the extinction $A_V$ up to 20 mag and more.

More than 40 young massive stars of spectral classes O--B3 and
supergiants, young open cluster NGC 1502 and 18 young stars of lower
masses give evidence of continuing star-forming process in the area.  A
revised list of members of the Cam OB1 association is compiled.  We
expect that special spectral and photometric observations must reveal
more T Tauri-type stars and related objects in the clouds.  In the
densest parts of the clouds more pre-stellar objects of large masses are
expected to be present.  The presence of a young stellar object GL\,490
in the densest part of the Tokyo 942 cloud may be considered as evidence
of continuing formation of massive stars.  In the next paper we will
describe more possible pre-stellar objects discovered by applying the
data from 2MASS, IRAS and MSX infrared surveys.

\vskip3mm

ACKNOWLEDGMENTS.  We are thankful to A. M. Dame for the CO radial
velocity data, to D. Froebrich for the dust distribution data, to G.
Gahm for the list of stars with H$\alpha$ emission, to B. Reipurth for
information on the new studies in our Milky Way segment, to
A.\,G. Davis Philip and S. Barta\v si\= ut\.e for helpful comments and
to E. Mei\v stas for help preparing the maps.  The use of the DSS,
2MASS, SkyView, WEBDA and Simbad databases is acknowledged.

\References

\refb Alvarez C., Hoare M., Glindemann A., Richichi A. 2004, A\&A, 427,
505

\refb Ann H. B., Lee S. H., Sung H. et al. 2002, AJ, 123, 905

\refb Barnard E. E. 1927, {\it Catalogue of 349 Dark Objects in the
Sky. A Photographic Atlas of Selected Regions of the Milky Way},
Carnegie Institution, Washington, D.C. = CDS catalogue VII/220A

\refb Brunt C. M., Kerton C. R., Pomerleau C. 2003, ApJS, 144, 47

\refb Campbell B., Persson S. E., McGregor P. J. 1986, ApJ, 305,
336

\refb Clemens D. P., Barvainis R. 1988, ApJS, 68, 257

\refb Dame T. M., Ungerechts H., Cohen R. S., de Geus E. J., Grenier I.
A., May J., Murphy D. C., Nyman L.-\AA., Thaddeus P. 1987, ApJ, 322,
706

\refb Dame T. M., Hartmann D.,  Thaddeus P. 2001, ApJ, 547, 792

\refb Digel S. W., Lyder D. A., Philbrick A. J., Puche D.,
Thaddeus P. 1996, ApJ, 458, 561

\refb Dobashi K., Uehara H., Kandori R., Sakurai T., Kaiden M.,
Umemoto T., Sato F. 2005, PASJ, 57, S1

\refb Elias P., Cabrera-Cano J., Alfaro E. J. 2006, AJ, 131, 2700

\refb Fich M., Blitz L. 1984, ApJ, 279, 125

\refb Finkbeiner D. P. 2003, {\it A Full-Sky H-alpha Template for
Microwave Foreground Prediction}, arXiv:astro-ph/0301558 v1

\refb Froebrich D., Murphy G. C., Smith M. D., Walsh J., Del Burgo C.
2007, MNRAS, 378, 1447

\refb Gahm G. 1990, personal communication

\refb Heithausen A., Stacy J. G., de Vries H. W., Mebold U., Thaddeus P.
1993, A\&A, 268, 265

\refb Herbig G. H., Bell K. R. 1988, Lick Obs. Bull., No. 1111

\refb Humphreys R. M. 1978, ApJS, 38, 309

\refb Humphreys R. M., McElroy D. B. 1984, {\it Catalogue of Stars in
Stellar Associations and Young Clusters}, Univ. of Minnesota, CDS
Strasbourg, Catalogue V/44

\refb Kharchenko N. V., Piskunov A. E., R\"oser S., Schilbach E.,
Scholz R.-D. 2005, A\&A, 438, 1163

\refb Kohoutek L., Wehmeyer R. 1997, {\it H-alpha stars in Northern
Milky Way}, Abh.  Hamburger Sternw. 11, Teil 1+2 = CDS III/205

\refb Launhardt R., Henning Th. 1997, A\&A, 326, 329

\refb Launhardt R., Sargent A. I. 2001, ApJ, 562, L173

\refb Lyder D. A. 2001, AJ, 122, 2634


\refb Maiz-Apell\'aniz J.,  Walborn N. R., Galu\'e H. A., Wei L. H.
2004, ApJS, 151, 103

\refb Melnik A. M., Efremov Yu. N. 1995, Astron. Letters, 21, 10

\refb Meyer M. R., Calvert N., Hillenbrand L. A. 1997, AJ, 114, 288

\refb Moffat A.\,F.\,J., Vogt N. 1973, A\&AS, 11, 3

\refb Ogura K., Sugitani K., Pickles A. 2002, AJ, 123, 2597

\refb Pe\~na J. H., Peniche R. 1994, Rev. Mex. AA, 28, 139

\refb Racine R. 1968, AJ, 73, 233

\refb Ruprecht J. 1964, {\it Revised List of OB Associations}, Trans.
IAU, 12B, 348

\refb Rydgren A. E., Vrba F. J. 1981, AJ, 86, 1069

\refb Samus N. N., Durlevich O. V. et al. 2004, {\it Combined General
Catalogue of Variable Stars}, Sternberg Astron. Inst., Moscow Univ. =
CDS II/250

\refb Schlegel D. J., Finkbeiner D. P., Davis M. 1998, ApJ, 500, 525

\refb Schwartz P. R. 1987, ApJ, 320, 258

\refb Snell R. L., Scoville N. Z., Sanders D. B., Erickson N. R.
1984, ApJ, 284, 176

\refb Stecklum B., Melnikov S. Y., Meusinger H. 2007, A\&A, 463, 621

\refb Strai\v{z}ys V. 1992, {\it Multicolor Stellar Photometry}, Pachart
Publ.  House, Tucson, Arizona

\refb Strai\v{z}ys V., Laugalys V. 2007, in {\it Handbook of Star
Forming Regions}, vol.\,1, ed. B. Reipurth, ASP Conf. Ser. (in press)

\refb Tapia M., Costero R., Echevarria J., Roth M. 1991, MNRAS, 253, 649

\refb Turner D. G., Evans N. R. 1984, ApJ, 283, 254

\refb Zdanavi\v cius J., Zdanavi\v cius K.,  Kazlauskas A. 1996,
Baltic Astronomy, 5, 563

\refb Zdanavi\v cius J., \v Cernis K., Zdanavi\v cius K., Strai\v zys V.
2001, Baltic Astronomy, 10, 349

\refb Zdanavi\v cius J., Zdanavi\v cius K. 2002a, Baltic Astronomy,
11, 75

\refb Zdanavi\v cius J., Zdanavi\v cius K. 2002b, Baltic Astronomy,
11, 441

\refb Zdanavi\v cius J., Strai\v zys V., Corbally C. J. 2002c,
A\&A, 392, 295

\refb Zdanavi\v cius J., Zdanavi\v cius K. 2005a, Baltic Astronomy,
14, 1

\refb Zdanavi\v cius J., Zdanavi\v cius K.,  Strai\v zys V. 2005b,
Baltic Astronomy, 14, 31

\refb Zeeuw P.\,T. de, Hoogerwerf R., Bruijne J.\,H.\,J. de, Brown
A.\,G.\,A., Blaauw A. 1999, AJ, 117, 354

\vskip4mm

{\bf Errata.} In the arXiv version of the paper the following errors
found in the printed version are corrected: (1) In Figure 2 the cloud
number 1004 is changed to 1014; (2) In the text and in Table 3 the star
IRAS 0313+599 is changed to IRAS 03134+5958.

\end{document}